# Growth mode control of the free carrier density in SrTiO$_{3-\delta}$ films

A. Ohtomo[1,2] and H. Y. Hwang[1,3]

[1]*Bell Laboratories, Lucent Technologies, Murray Hill, New Jersey 07974, USA*
[2]*Institute for Materials Research, Tohoku University, Sendai, 980-8577, Japan*
[3]*Department of Advanced Materials Science, University of Tokyo, Kashiwa, 277-8651, Japan*

(Dated: April 5, 2006)

We have studied the growth dynamics and electronic properties of SrTiO$_{3-\delta}$ homoepitaxial films by pulsed laser deposition. We find the two dominant factors determining the growth mode are the kinetics of surface crystallization and oxidation. When matched, persistent two-dimensional layer-by-layer growth can be obtained for hundreds of unit cells. By tuning these kinetic factors, oxygen vacancies can be frozen in the film, allowing controlled, systematic doping across a metal-insulator transition. Metallic films can be grown, exhibiting Hall mobilities as high as 25,000 cm$^2$/Vs.

## I. INTRODUCTION

Aspects of SrTiO$_3$ have been studied for many years, including the anomalous low-temperature dielectric response due to incipient ferroelectricity[1], and electronic properties upon doping[2-6]. Oxygen vacancies induce free electrons with unusually high mobility for such a narrow-band system, with significant charged-impurity screening provided by the lattice. Remarkably, the evolution from a dielectric insulator, to a doped semiconductor, then to a metal and superconductor all occurs within the first 0.03 % of oxygen vacancies. Recently, thin film SrTiO$_3$ has been actively investigated to utilize this wide range of physical properties. Motivations include functions such as high-K dielectric layers for advanced CMOS[7], insulating tunnel barriers[8], semiconductor for oxide field effect devices[9], conducting layers[10], etc. In addition, SrTiO$_3$ is a prominent epitaxial substrate for many of the perovskite-derived transition metal oxides.

In all of these contexts, SrTiO$_3$ homoepitaxy provides an ideal prototype to evaluate the growth dynamics in the current efforts to develop complex oxide heteroepitaxial structures. *In-situ* reflection high-energy electron diffraction (RHEED) techniques have been successfully applied to demonstrate near-perfect epitaxy. Layer-by-layer growth sustained by growth interruptions[11], as well as step-flow growth[12], has been demonstrated using pulsed-laser deposition (PLD). However, a general understanding of the growth dynamics, including the resulting oxygen stoichiometry, has not yet been established. This is a key issue underlying applications of thin film SrTiO$_3$, as well as an opportunity to provide controlled dopant profiles in heteroepitaxial structures.

Here we report our study of the growth dynamics of SrTiO$_{3-\delta}$ homoepitaxial films using PLD. By varying the temperature ($T_g$) and the oxygen partial pressure ($P_{O2}$) during growth, we have mapped out the growth mode transition from three-dimensional island growth to two-dimensional layer-by-layer growth. The optimal conditions for extended layer-by-layer growth occur when the time constant for crystallization ($\tau_{cryst}$) of the adatom species matches the surface oxidation time constant ($\tau_{ox}$). This also marks a key threshold for the electronic properties of the films, because when $\tau_{cryst} < \tau_{ox}$, oxygen vacancies are frozen in the growing film, resulting in free electron carriers. The films can be systematically doped in this manner to undergo a metal-insulator transition, with low temperature mobilities as high as 25,000 cm$^2$/Vs.

## II. EXPERIMENT

Homoepitaxial SrTiO$_3$ films 1000 Å thick were grown by PLD from a single crystal target on atomically flat, TiO$_2$-terminated (001) SrTiO$_3$ substrates. The entire growth process was monitored by RHEED. Substrates were mounted using silver paint or clamping on an oxidized inconel surface, and $T_g$ was measured by a pyrometer assuming as emissivity of 0.8. $T_g$ and $P_{O2}$ were varied in this study, while the repetition rate (5 Hz) and fluence (~ 1 J/cm$^2$) of the KrF excimer laser were held constant. After growth, the temperature was lowered to room temperature at a constant rate of 50 °C/minute while keeping $P_{O2}$ constant. The obtained films were characterized by means of atomic force microscopy (AFM) and x-ray diffraction (XRD) at the National Synchrotron Light Source at Brookhaven National laboratory as well as using a laboratory apparatus. Transport properties were measured in six-probe Hall bar geometries using evaporated Al Ohmic contacts. Films were also grown on (001) LaAlO$_3$ substrates in order to measure their compositions using Rutherford backscattering spectroscopy (RBS).

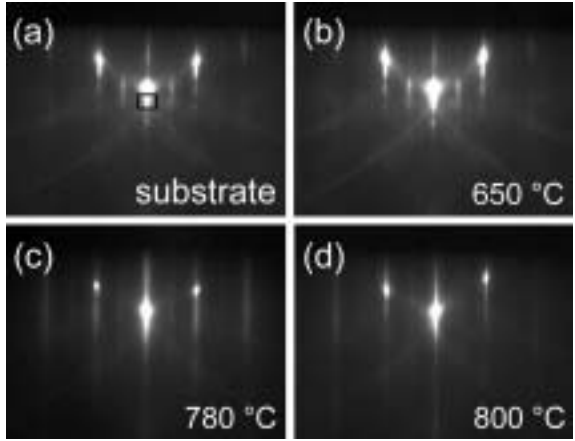

FIG. 1. RHEED patterns of SrTiO$_3$ substrate (a) and homoepitaxial films 1000 Å thick grown at 650 °C, 780 °C and 800 °C (b-d). The images were taken along the [100] azimuth with 25 keV electrons. The incidence angle was 1.1° (in-phase condition). The measurement temperatures were same as the growth temperatures. The intensity of specular spot outside the Laue circle, indicated by the square in (a), was monitored to record RHEED intensity oscillations shown in Fig. 2.

## III. RESULTS AND DISCUSSION

In order to analyze the growth-induced effect in RHEED, we set a constant diffraction condition through all deposition experiments – the in-phase condition where the Bragg condition is satisfied for the SrTiO$_3$ unit cell. The fine diffraction spots were observed from the substrates after heating at 950 °C in $P_{O2} = 1\times10^{-6}$ Torr for 30 min, as shown in Fig. 1(a). Weak streaks between the specular and (01) spots are presumably due to the (2x1) reconstruction[13]. These diffraction features hardly changed during homoepitaxial growth, when grown at 650 °C in $P_{O2} = 1\times10^{-6}$ Torr [Fig. 1(b)]. The films grown at higher $T_g$ showed streaky patterns along the streak direction as shown in Fig. 1(c) and 1(d). Figure 2 shows the normalized RHEED intensity oscillation of the specular spot during the growth in $P_{O2} = 1\times10^{-6}$ Torr at various temperatures. Each oscillation corresponds to the completion of one perovskite unit cell, and the typical growth rate was one unit cell for ~ 30 laser pulses. The topmost trace ($T_g = 650$ °C) represents the optimal conditions for this $P_{O2}$. After initial decay in amplitude, the RHEED oscillations persist in steady state after more than 100 unit cells of growth. Many non-intrinsic factors can dampen RHEED oscillations[14], thus this trace confirms our efforts to optimize the experimental geometry – minimizing flux in-homogeneity, for example. Any observed factor contributing to oscillation persistence, therefore, can be considered *intrinsic* and growth mode related. In fact, almost same oscillation traces were also obtained from the (01) spots. For higher and lower $T_g$, both the oscillation amplitude and average intensity decrease more quickly (higher $T_g$ shown in Fig. 2) and eventually the patterns become diffused [Fig. 1(c) and 1(d)].

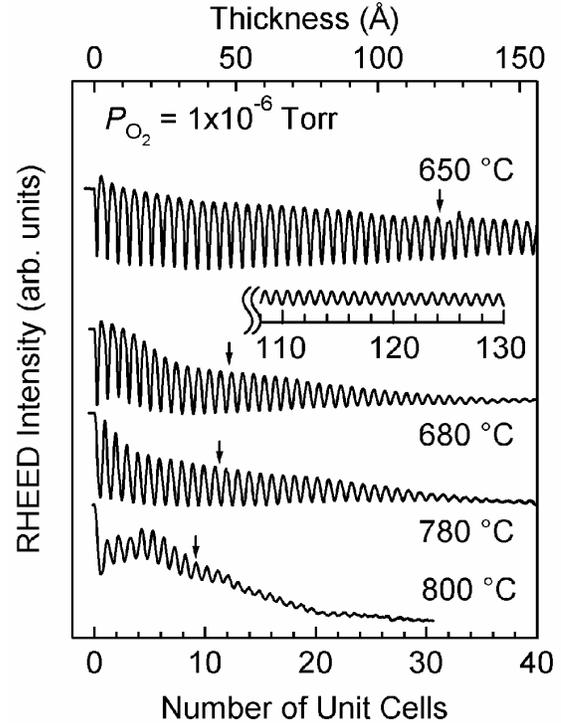

FIG. 2. RHEED intensity oscillations during SrTiO$_{3-\delta}$ (001) homoepitaxial growth at various temperatures in an oxygen partial pressure $P_{O2} = 10^{-6}$ Torr. Arrows indicate the position of the loss of half the initial oscillation amplitude.

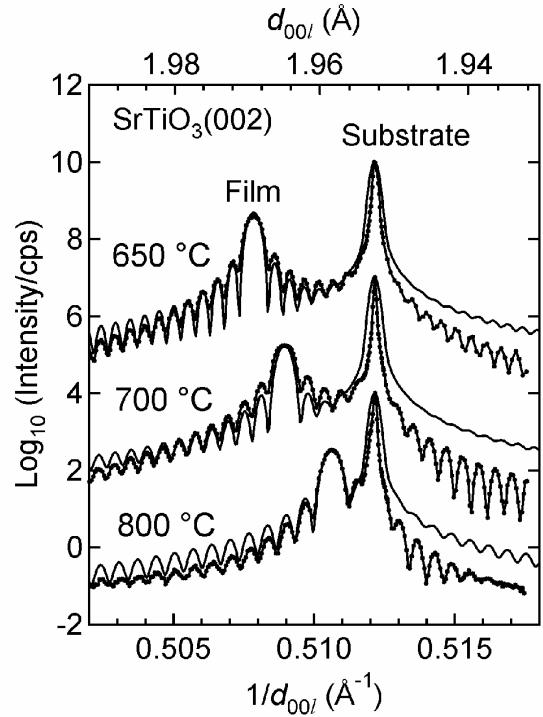

FIG. 3. $\theta - 2\theta$ x-ray diffraction patterns (dots) and simulation curves (solid lines) near the (002) Bragg condition for three films grown in $P_{O2} = 10^{-6}$ Torr (650 °C and 700 °C) and $10^{-5}$ Torr (800 °C).

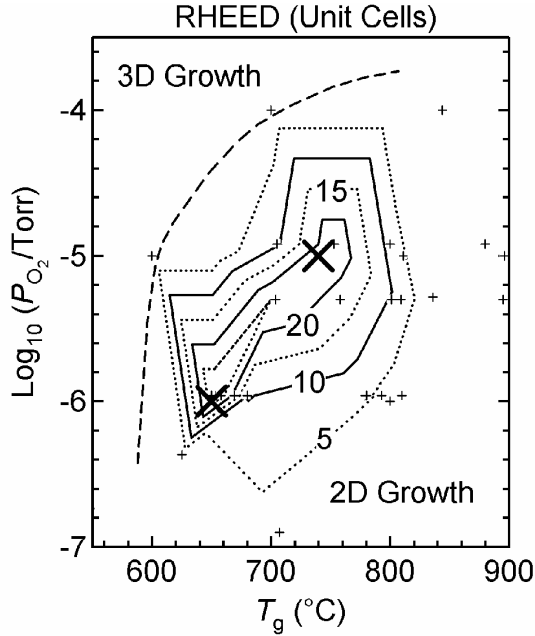 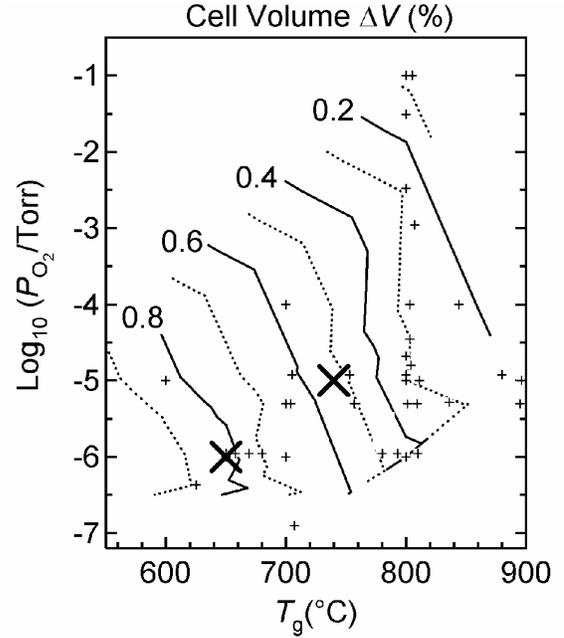

FIG. 4. SrTiO$_{3-\delta}$ (001) homoepitaxial growth phase diagram as a function of $P_{O2}$ and growth temperature ($T_g$). Contour lines denote the number of unit cells for which half the RHEED oscillation amplitude is lost. The thick dashed line indicates the boundary between two-dimensional layer-by-layer growth and three-dimensional island growth modes. Thick crosses depict the optimal conditions for RHEED persistence at $P_{O2} = 10^{-5}$ and $10^{-6}$ Torr.

FIG. 5. Coutour map of the increase in unit cell volume, arising from increase in the film c-axis as a function of $P_{O2}$ and $T_g$. Thick crosses depict the optimal conditions for RHEED persistence at $P_{O2} = 10^{-5}$ and $10^{-6}$ Torr.

Figure 3 shows the $\theta - 2\theta$ synchrotron XRD pattern near the (002) Bragg condition for three films grown in $P_{O2} = 1 \times 10^{-6}$ and $1 \times 10^{-5}$ Torr on SrTiO$_3$. Although this is homoepitaxial growth, the film peak is clearly separated from the substrate, indicating elongation along the c-axis. Off-axis reciprocal space mapping in a laboratory XRD confirms that all of the films in this study are fully strained. Therefore, the longer c-axis corresponds to an expanded lattice volume change ($\Delta V$) of 0.85, 0.63, and 0.30 % for the films grown at 650, 700, and 800 °C, respectively.

The results spanning a wide range of parameters are summarized in the growth phase diagram of Fig. 4. Contour lines (thin solid and dashed lines) denote the number of unit cells of growth for which the RHEED oscillation loses half of its initial amplitude, indicated by arrows in Fig. 2. The optimal condition for extended layer-by-layer growth is oriented diagonally, with optimal conditions at $P_{O2} = 1 \times 10^{-5}$ and $1 \times 10^{-6}$ Torr marked by crosses. At higher $P_{O2}$ and lower $T_g$, the lower surface mobility of the adatoms leads to island growth contributions that diminish the RHEED intensity. Consistent with this, the thick dashed line gives the boundary for surface roughening, as observed by *ex-situ* atomic force microscopy. Outside of this region, the atomic step and terrace structure of the original substrate surface is preserved on the surface of the film. Figure 5 gives the phase diagram of the unit cell volume expansion as a function of growth conditions, taken from a mix of synchrotron and laboratory x-ray sources[15]. This increase in $\Delta V$, arising from the expanded c-axis, is found to increase monotonically with decreasing $P_{O2}$ and lowering $T_g$.

In order to understand the structure of Fig. 4, many dynamical factors should be considered, two of which dominate the growth phase diagram. Among the various thin film growth methods commonly used, PLD is generally the most kinetic, particularly at low chamber pressures where the mean free path of the ablated species is considerably longer than the target-substrate distance. This results in unusually high surface diffusion energy. The crystallization of the surface precursors is kinetically limited with activation barriers for the surface diffusion and dissociation of small islands. At very low $T_g$, the adatoms freeze in metastable form, resulting in an extremely long $\tau_{cryst}$. With increasing $T_g$, $\tau_{cryst}$ decreases due to thermal assistance. The temporal evolution of the RHEED intensity recovery after growth interruptions can be used to estimate $\tau_{cryst}$, and it has been found to follow the Arrhenius form $\tau_{cryst}^{-1} = A\exp(-E_{Acryst}/k_B T_g)$, where A is a pre-exponential factor, $E_{Acryst}$ is the activation energy, and $k_B$ is the Boltzmann constant[16]. $E_{Acryst}$ is reported to be ~3.8 eV, although increasing the effective nuclei and step edge density reduces this value[17, 18].

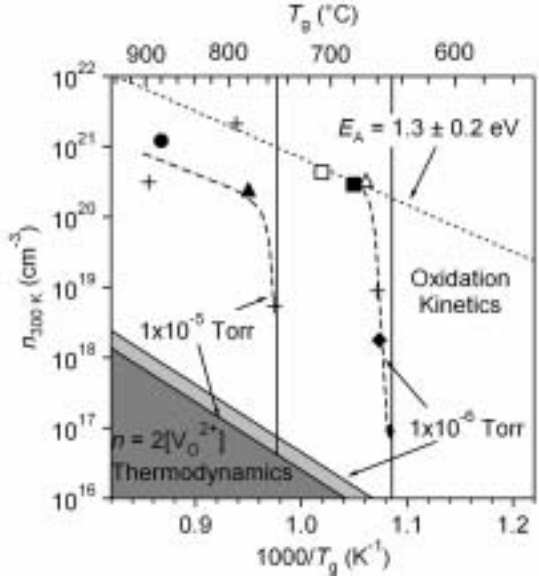

FIG. 6. Variation of the room temperature free carrier density $n_{300\,K}$ with $T_g$ at $P_{O2} = 10^{-5}$ and $10^{-6}$ Torr. With decreasing $T_g$, $n_{300\,K}$ suddenly drops and extrapolates to the carrier densities corresponding to thermodynamic equilibrium oxygen vacancies, indicated for $P_{O2} = 10^{-5}$ and $10^{-6}$ Torr. This point (marked by vertical lines), also corresponds to the optimal conditions for RHEED persistence (Fig. 4).

The other important kinetic limit is surface oxidation. Since repeated laser ablation results in an oxygen deficient surface on the target oxide[19], reduced species on the film surface are oxidized during surface migration and after crystallization. This has been studied by *in-situ* optical reflectivity measurements during PLD growth of *monolayer* $SrTiO_3$[20]. Here also an Arrhenius form was observed, where $E_{Aox}$ is found to be $P_{O2}$ independent at ~ 1.35 eV, and $A = 4 \times 10^9 P_{O2}$, the pre-exponential factor, captures the $P_{O2}$ dependence. Taken together, the dynamics of crystallization and oxidation give a framework to Fig. 4. Although both follow the same simple rate equation, the different constants result in the trend that with growth conditions varying from high $P_{O2}$-low $T_g$ to low $P_{O2}$-high $T_g$, there is a crossover from $\tau_{cryst} > \tau_{ox}$ to $\tau_{cryst} < \tau_{ox}$. Optimal layer-by-layer growth is realized on the verge of three-dimensional growth when $\tau_{cryst} \sim \tau_{ox}$. The conditions with higher $T_g$ and lower $P_{O2}$ also hamper layer-by-layer growth, not due to the partial contribution of step-flow-growth, which occurs either at much higher $T_g$ or for low miscut substrates[17], but due to mismatching kinetic factors $\tau_{cryst} < \tau_{ox}$, resulting in an irregular nucleation and growth cycle. In this region, the RHEED intensity recovered on the minute timescale at a growth interruption.

It is relevant to compare these time constants with the inverse repetition rate of the laser pulses. Both $\tau_{cryst}$ and $\tau_{ox}$ just after monolayer growth at $P_{O2} = 1 \times 10^{-5}$ Torr and $T_g = 700\ °C$ are more than $10^2$ seconds[17,20], which is much longer than the laser pulse interval (0.2 seconds).

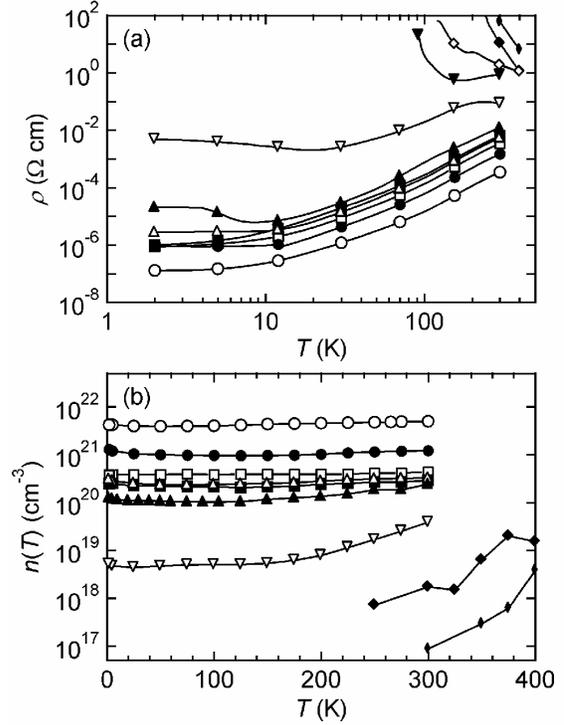

FIG. 7. (a) Temperature dependent resistivity for $SrTiO_{3-\delta}$ films varying $\delta$ following Fig. 6. (b) $n(T)$ corresponding to the films in panel (a). Different symbols for each trace indicate correspondence to the data shown in Fig. 6 and Fig. 8.

Applying a longer pulse interval, for further evolution of the surface kinetics, is not practical. Indeed, no change in RHEED oscillation persistence is observed by decreasing the laser repetition from 5 to 0.5 Hz. It should be mentioned that the growth mode can be manipulated by changing other parameters such as the laser energy and the chemical potential of the source materials (ambient gases and target solids). Many dynamical factors are involved and interdependent in a manner not discussed above, where we have focused on the key parameters in typical growth conditions.

In addition to controlling the growth modes, the kinetics of crystallization and oxidation dominate the electrical properties of the films as well. For $T_g$ above that for optimal layer-by-layer growth ($\tau_{cryst} < \tau_{ox}$), excess oxygen vacancies beyond thermodynamic equilibrium are frozen in the growing film. Because doubly-charged oxygen vacancies ($V_O^{2+}$) in $SrTiO_3$ act as donors, contributing two electrons each, they lead to a dramatic increase in the free carrier density[21]. Figure 6 shows a central result of this study: *free carriers suddenly appear as $\tau_{cryst} < \tau_{ox}$, and the resulting carrier density reflects the competing rates for crystallization and oxidation*. The room temperature carrier density $n = -1/eR_H$, where $R_H$ is the Hall coefficient and $e$ is the electron charge, is shown for varying $T_g$ at $P_{O2} = 1 \times 10^{-5}$ and $1 \times 10^{-6}$ Torr. Clearly seen are dramatic increases of $n$ by more than three orders of magnitude, just at the optimal $T_g$ for extended layer-by-layer growth where $\tau_{cryst} \sim \tau_{ox}$ (crosses in Figs. 4 and 5).

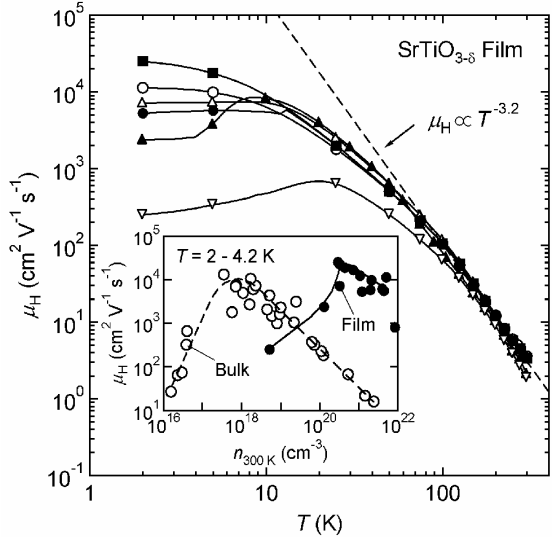
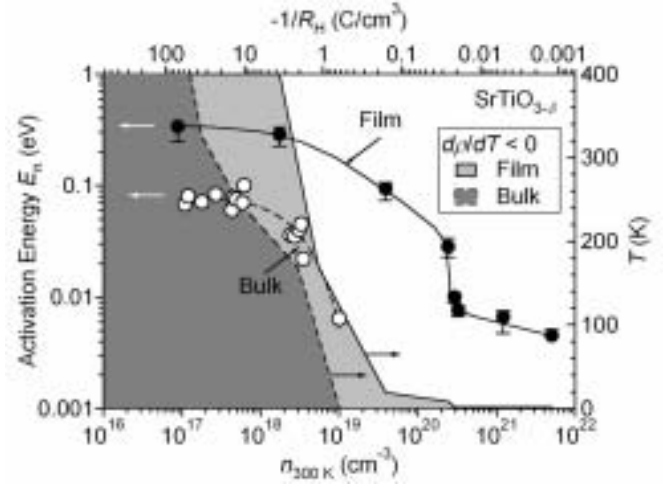

FIG. 8. Temperature dependence of the Hall mobility for metallic SrTiO$_{3-\delta}$ films. The dashed line gives the best power law fit between 100 and 300 K. The symbols correspond to those in Fig. 7. Inset: Low temperature Hall mobility as a function of carrier density for the thin films of this study and bulk single crystals (taken from Refs. 3-6, 24).

FIG. 9. Carrier activation energy as a function of $n_{300 K}$ for thin films and bulk single crystals (taken from Refs. 5, 6). The temperature dependent boundary for the metal-insulator transition (d$\rho$/d$T$ < 0) for thin films and bulk crystals is indicated by light and dark grey, respectively.

At lower $T_g$, the falling $n$ values extrapolate to the thermodynamic equilibrium V$_O^{2+}$ densities $n = 2[V_O^{2+}]$ expected from simple free energy considerations[22], as confirmed for bulk single crystals[23, 24]. The activation energy for the $T_g$ dependence of $n$ at high temperatures is found to be 1.3 ± 0.2 eV, in good agreement with $E_{Aox}$ discussed previously[20]. The highest electron densities observed is ~ 5x10$^{21}$ cm$^{-3}$, suggesting that ~ 5 % of the oxygen sites are unoccupied (a chemical formula of SrTiO$_{2.85}$). It is not surprising that such grossly oxygen deficient films can still maintain the perovskite lattice, given the known stability of SrTiO$_{3-\delta}$ up to $\delta$ = 0.28 (Ref. 24). RBS studies of films grown on LaAlO$_3$ confirm the cation stoichiometry Sr$_1$:Ti$_1$ to within 1%, eliminating this possibility as a major contribution to the free carrier density.

Figures 7(a) and 7(b) show the temperature dependent resistivity $\rho(T)$ and $n(T)$ for conducting samples. The systematic change from metallic to insulating films corresponds to a decrease in $n$, and the metal insulator transition occurs near $n = 10^{19}$ cm$^{-3}$. Highly conducting samples show nearly temperature independent $n$, while insulating samples show activation behavior with decreasing $n$. In the films studied here, no systematic differences in $\rho(T)$ were found for equivalent $n$ achieved by different $T_g$ and $P_{O2}$ during growth. Furthermore, the high kinetic barrier to oxidation and the slow oxygen diffusion times at the growth parameters allow the identification of the measured transport properties with the frozen configuration due to the kinetics of growth.

The temperature dependent Hall mobility $\mu_H(T)$ for the metallic samples in Fig. 7 is plotted in Fig. 8. At higher temperatures, all of the data follow essentially the same curve with $\mu_H \sim T^{-3.2}$, very similar to previous reports for bulk single crystal samples of SrTiO$_{3-\delta}$ (Ref. 2). At low temperatures, the maximum $\mu_H$ = 25,000 cm$^2$/Vs was observed for a film with $n = 2.5$x$10^{20}$ cm$^{-3}$. To our knowledge, this is the highest reported $\mu_H$ for doped SrTiO$_3$ in any form. However, despite the similarity of $\mu_H(T)$ in our films with bulk single crystals, there are several significant differences that bear consideration. The inset to Fig. 8 displays the carrier density dependence of $\mu_H$(2 - 4.2 K) for both previous bulk work and the films of this study. Although both share a peak in mobility of comparable magnitude and form, this peak occurs at much higher carrier density in the films than in bulk.

Some insight to the difference between thin film and bulk results can be found in a comparison of the evolution of the carrier activation energy $E_n$ with carrier density, shown in Fig. 9. A dramatic increase in $E_n$ is observed at a much higher carrier density in the films as compared to bulk single crystals. One contribution to this shift may arise from the recently observed effects of surface depletion[25]. A second contribution is the diminished low temperature dielectric constant $\varepsilon$ often observed for thin films, noting that $E_n \sim \varepsilon^{-2}$ for a hydrogen-like donor. Typical low temperature values for films range from 500 – 1500, whereas the single crystal value exceeds 20,000 (Ref. 26). This difference is likely closely related to the observed volume expansions (Fig. 5). The tendency for a tetragonal volume expansion has been observed in bulk quenched SrTiO$_{3-\delta}$, but it is greatly diminished for slow-cooled samples ($\Delta V \sim 0.2$ % for SrTiO$_{2.72}$) (Ref. 24).

We note that by moderate oxygen post annealing (400 °C, 2 hours in 1 atm of $O_2$), the metallic films become completely insulating, while maintaining the expanded lattice. The volume expansion has also been observed in much higher $P_{O2}$ during PLD growth[27]. Taken together, these results indicate a nontrivial modification of the lattice arising from the non-equilibrium kinetics of PLD growth.

## IV. CONCLUSIONS

Using *in-situ* RHEED, we have carefully examined the evolution of the PLD growth mode during $SrTiO_3$ homoepitaxy, which strongly depends on the oxygen partial pressure and growth temperature. Nearly ideal oxygen stoichiometry and extended layer-by-layer growth were simultaneously achieved under conditions where the time constants of two key kinetic processes, oxidation and crystallization, match. By detuning from this condition, the oxygen stroichiometry could be systematically controlled in the growing film. Our approach provides a way to vary oxygen vacancy profiles in complex oxide heterostructures while keeping atomically sharp heterointerfaces[10,28].

We have further investigated the structural and electronic properties of homoepitaxial $SrTiO_{3-\delta}$ films over a wide range of oxygen $\delta$. Films with high $\delta$ exhibited metallic conduction down to low temperatures, and with decreasing $\delta$, the films undergo a metal-insulator transition around $\delta \sim 0.006$. The variation of the low temperature Hall mobility as a function of carrier density was found to be qualitatively similar to that of bulk single crystals, but with a maximum at much higher carrier density in the films than in bulk. This difference can be attributed to the observed tetragonal volume expansion causing reduction of the background dielectric constant far below the bulk crystalline value[29].

## ACKNOWLEDGMENTS

We thank K. Evans-Lutterodt for assistance with synchrotron XRD, and M. Siegert and D. C. Jacobson for assistance with RBS. A.O. acknowledges partial support by the Nishina Memorial Foundation, Japan.

*Corresponding author. Electronic address: hyhwang@k.u-tokyo.ac.jp
[1] K. A. Müller and H. Burkard, Phys. Rev. B **19**, 3593 (1979).
[2] H. P. R. Frederikse, W. R. Thurber, and W. R. Hosler, Phys. Rev. **134**, A442 (1964).
[3] O. N. Tufte and P. W. Chapman, Phys. Rev. **155**, 796 (1967).
[4] H. P. R. Frederikse and W. R. Hosler, Phys. Rev. **161**, 822 (1967).
[5] C. Lee, J. Yahia, and J. L. Brebner, Phys. Rev. B **3**, 2525 (1971).
[6] C. Lee, J. Destry, and J. L. Brebner, Phys. Rev. B **11**, 2299 (1975).
[7] R. A. McKee, F. J. Walker, and M. F. Chisholm, Science **293**, 468 (2001).
[8] J. M. De Teresa, A. Barthelemy, A. Fert, J. P. Contour, R. Lyonnet, F. Montaigne, P. Seneor, and A. Vaures, Phys. Rev. Lett. **82**, 4288 (1999).
[9] K. Ueno, I. H. Inoue, H. Akoh, M. Kawasaki, Y. Tokura, and H. Takagi, Appl. Phys. Lett. **83**, 1755 (2003).
[10] A. Ohtomo, D. A. Muller, J. L. Grazul, and H. Y. Hwang, Nature **419**, 378 (2002).
[11] D. H. A. Blank, G. Koster, G. Rijnders, E. van Setten, P. Slycke, and H. Rogalla, Appl. Phys. A **69**, S17 (1999).
[12] M. Lippmaa, N. Nakagawa, M. Kawasaki, S. Ohashi, Y. Inaguma, M. Itoh, and H. Koinuma, Appl. Phys. Lett. **74**, 3543 (1999).
[13] M. R. Castell, Surf. Sci. **505**, 1 (2002).
[14] W. Braun, *Applied RHEED*, vol. 154 of *Springer Tracts in Modern Physics* (Springer-Verlag, Berlin, 1999).
[15] Films grown above $10^{-4}$ Torr include Nb and La doped $SrTiO_3$.
[16] H. Karl and B. Stritzker, Phys. Rev. Lett. **69**, 2939 (1992).
[17] M. Lippmaa, N. Nakagawa, M. Kawasaki, S. Ohashi, and H. Koinuma, Appl. Phys. Lett. **76**, 2439 (2000).
[18] J. H. Neave, P. J. Dobson, B. A. Joyce, and J. Zhang, Appl. Phys. Lett. **47**, 100 (1985).
[19] L. M. Doeswijk, G. Rijnders, and D.H.A. Blank, Appl. Phys. A **78**, 263 (2004).
[20] X. D. Zhu, W. Si, X. X. Xi, and Q. Jiang, Appl. Phys. Lett. **78**, 460 (2001).
[21] At higher oxygen vacancy concentrations, $[V_O^{2+}] > 5 \times 10^{20}$ cm$^{-3}$, the numbers of free electrons estimated from the Hall effect may be smaller than $2[V_O^{2+}]$ (see Ref. 24).
[22] N.-H. Chan, R. K. Sharma, and D. M. Smyth, J. Electrochem. Soc. **128**, 1762 (1981).
[23] H. Yamada and G. R. Miller, J. Solid State Chem. **6**, 169 (1973).
[24] W. Gong, H. Yun, Y. B. Ning, J. E. Greedan, W. R. Datars, and C. V. Stager, J. Solid State Chem. **90**, 320 (1991).
[25] A. Ohtomo and H. Y. Hwang, Appl. Phys. Lett. **84**, 1716 (2004).
[26] T. Sakudo and H. Unoki, Phys. Rev. Lett. **26**, 851 (1971).
[27] E. J. Tarsa, E. A. Hachfeld, F. T. Quinlan, J. S. Speck, and M. Eddy, Appl. Phys. Lett. **68**, 490 (1996).
[28] D. A. Muller, N. Nakagawa, A. Ohtomo, J. L. Grazul, and H. Y. Hwang, Nature **430**, 657 (2004).
[29] A. A. Sirenko, I. A. Akimov, J. R. Fox, A. M. Clark, H-C. Li, W. Si, and X. X. Xi, Phys. Rev. Lett. **82**, 4500 (1999).